\documentclass[%
 reprint,
superscriptaddress,
 amsmath,amssymb,
aps,prl
]{revtex4-2}

\usepackage{graphicx}% Include figure files
\usepackage{dcolumn}% Align table columns on decimal point
\usepackage{bm}% bold math
\usepackage{color}
\usepackage{multirow}

\begin{document}

\preprint{APS/123-QED}

\title{On how walls shape dissipation intermittency}

\author{Peng-Yu Duan}
\author{Xi Chen}
\email{chenxi97@outlook.com}
\affiliation{
 Key Laboratory of Fluid Mechanics of Ministry of Education, Beihang University (Beijing University of Aeronautics and Astronautics), Beijing 100191, PR China
}
\author{Katepalli R. Sreenivasan}
\email{krs3@nyu.edu}
\affiliation{%
 Tandon School of Engineering, Courant Institute of Mathematical Sciences, and Department of Physics, New York University, New York 10012, USA
}%

\date{\today}% It is always \today, today,

\begin{abstract}
Intermittency of energy dissipation has long been studied via high-order moments in homogeneous and isotropic turbulence (HIT), but not much where the boundary effects are explicitly included. Here, we derive two fundamental Reynolds number scaling expressions for dissipation moments in wall-bounded flows---one in the outer region where the boundary effects are weak and the other close to the walls where those effects are strong---and support these expressions by direct numerical simulations. Dissipation moments in the outer region follow universal power laws with exponents linked to anomalous scaling of velocity structure functions in HIT. In contrast, moments near the wall follow a bounded defect law, leading to a finite asymptotic limit without intermittency.
For very large Reynolds numbers, the outer proposal predicts vanishing dissipation compared to that on the wall, highlighting
the need for solid boundaries in generating Onsager-type singularities.

\end{abstract}

\maketitle

\textit{Introduction---}Turbulence is ubiquitous and constantly challenging, thus inspiring innovative efforts to describe, explore, and understand its complex nature.
Far from boundaries and at sufficiently high Reynolds numbers, turbulence is expected to be essentially homogeneous and isotropic \cite{Frisch_turb,Sreeni1997_ARFM_turb}. In this region, the interscale energy transfer process dominates. In three-dimensional turbulence, the caricature of this process is a cascade of energy from the integral scale $L$ to the dissipation scale $\eta$ \cite{Richardson1922_cascade}, which was succinctly formulated by Kolmogorov in 1941 (K41 henceforth) \cite{K41}. In the inertial range ($\eta \ll r \ll L$) where the cascade is thought to be inviscid and self-similar, K41 predicts that the velocity structure functions are determined only by $\langle\epsilon\rangle$ and follow power law scaling $S_p(r) \equiv \langle (\delta_r u)^p \rangle \propto (\langle\epsilon\rangle r)^{p/3}$. Here, $\delta_r u = u(x+r) - u(x)$, the longitudinal increment of the velocity component $u$ (with $x$ and $r$ in same direction as $u$), carries information on energy transfer between eddies of scale $r$, $\epsilon$ is the local turbulent dissipation rate, and $\langle \cdot \rangle$ denotes an ensemble average.
Underlying K41 is the constancy of mean energy dissipation. %, which is also the main objection of Landau to K41's universality due to the intermittency of dissipation \cite{Landau1987_fluid}.
In the presence of intermittency for real turbulence, the scaling laws for structure functions differ from K41, require modifications (termed anomalous scaling) of the type $S_p(r) \sim r^{\zeta_p}$, with the exponents $\zeta_p$ depending nonlinearly on $p$, and must be determined separately for each $p$. Such corrections, together with the understanding of the failure of K41 and the prevalence of dissipation intermittency, have long been and remain a central topic in modern turbulence research \cite{Dubrulle2019_JFM_Perspective, Sreeni2025_ARFM_turb}.

For practical flows, such as channels, pipes, and flow over airfoils, the situation becomes more complex because of the wall. Recently, the universality of small scales in different flow geometries has been observed in terms of the Reynolds number dependence of dissipation \cite{Schumacher2014_PNAS_universality, Buaria&Pumir2025_PRF_extreme_events}, but these studies focus on the flow center, where a state similar to homogeneous and isotropic turbulence (HIT) develops naturally. In contrast, near the wall, strong mean shear breaks conditions of HIT, and small-scale turbulence feels directly the geometric constraints or large-scale features \cite{Luminita2002_PoF_small_scale_in_CH,Vassilicos2017_PoF_diss,DCS2025_JFM_circulation}, thus violating assumptions of local universality.
As an extreme example, velocities vanish identically at the wall while dissipation reaches its maximum and remains highly fluctuating. It is thus essential to examine how dissipation moments vary with wall-normal distance. However, the influence of the wall on the intermittency of small-scale turbulence has not been explored much. Rectifying this situation is our goal.

%The anomalous scaling of velocity structure functions is found to be related to the Reynolds number scaling of dissipation moments $\langle\epsilon^p\rangle$ for homogeneous and isotropic turbulence (HIT) \cite{Sreeni&Yakhot2021_PRF_scaling, Yakhot1998_PRE_pdf, Yakhot2006_pdf}.

%We hence focus on the Reynolds number scaling of dissipation fluctuations in turbulent channel flows, both near the wall and away from it, in wall-parallel planes where turbulence remains homogeneous in statistical sense.

To anticipate, we show here that turbulent dissipation moments in wall flows follow the universal power law in the outer region (including the log layer), given by
\begin{equation}
    \langle \epsilon_{}^{p} \rangle ^\frac{1}{p}/\epsilon_{\tau} =  C_p Re_\tau^{-1+\gamma_p},
\label{eq:diss_outer_power}
\end{equation}
while at the wall they follow the bounded defect law
\begin{equation}
    \langle \epsilon_{}^{p} \rangle ^\frac{1}{p}/\epsilon_{\tau} =  \alpha_p - \beta_p Re_\tau^{-1/4}.
\label{eq:diss_wall_defect}
\end{equation}
In the above, $\epsilon$ is the streamwise or spanwise dissipation (explicit expressions are given later), $\epsilon_{\tau}=u^4_\tau/\nu$ is the mean flow dissipation at the wall, $Re_\tau=u_\tau H/\nu$ is the friction Reynolds number (with $H$ the flow thickness, $u_\tau$ the friction velocity and $\nu$ the kinetic viscosity), and $\gamma_p$ is the intermittency correction in the outer flow ($\gamma_p = 0$ without intermittency); $C_p$, $\alpha_p$ and $\beta_p$ are coefficients independent of $Re_\tau$ but may depend on order $p$. The rest of the paper deals with the derivation of Eqs.~\eqref{eq:diss_outer_power} and \eqref{eq:diss_wall_defect} and their assessment by means of data, followed by a discussion of their significance.

%$\epsilon$ represents streamwise or spanwise dissipation (explicit expressions given later),

\begin{figure}
  \includegraphics[scale=1.0]{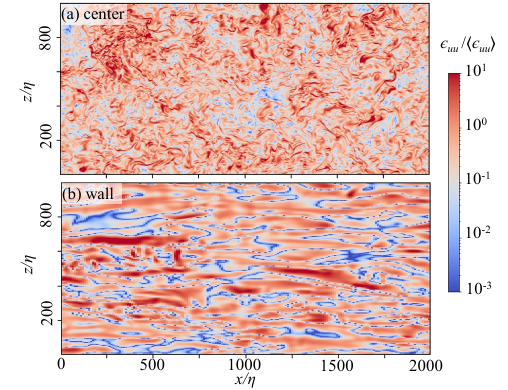}
 \caption{\label{fig:diss_snapshots} Typical snapshots of the streamwise component of turbulent dissipation rate on (a) the channel centerplane and (b) the wall, normalized by the ensemble-averaged component turbulent dissipation rate $\langle\epsilon_{uu}\rangle$, obtained from direct numerical simulations (DNS) of channel flow in Ref.~\citep{LM2015} at $Re_\tau=5200$.  Spatial domains are nondimensionalized using the Kolmogorov length $\eta$. }
\end{figure}

%\textit{Data.---}Direct numerical simulation data of turbulent channels for $Re_\tau=180$, 550, 1000, and 5200 are used to calculate high-order moments. Our data for the first two Reynolds numbers have been validated in Refs. \cite{Jiabin21, DCS2025_JFM_circulation}, while data for $Re_\tau=1000$ and 5200 are available from the Johns Hopkins Turbulence Database \citep{LM2015, JHU_database}. Details of the simulation settings are described in the Supplemental Material \cite{supply}.

% For large asymptotic $Re_\tau$, (1) implies a vanishing dissipation while (2) indicates a finite value of $\alpha_p$, thus marking a sharp difference between intermitence, consistent with our recent findings on wall-parallel circulations \citep{DCS2025_JFM_circulation}.

\textit{Patterns of dissipation ---} A direct illustration of the shear effect on dissipation is given in Fig.~\ref{fig:diss_snapshots}, which compares, for the channel, the patterns of the streamwise component of the dissipation rate $\epsilon$ ($\epsilon_{uu}=\nu\partial_j u \partial_j u$, where $u$ is the fluctuation velocity after subtracting the mean) on horizontal planes ($x$-$z$) at the channel center ($y/H=1$) and on the wall ($y/H=0$).  The two domain sizes are specified in Kolmogorov lengths, ensuring comparable information on dissipative scales.
On the centerplane, the dissipation field exhibits no preferred direction and shows a nearly isotropic character.
In contrast, dissipation at the wall forms a streaky structure in the streamwise direction (similar to other quantities there \citep{LM2015}), with fluctuations reaching more intense extrema.

\begin{figure}
\includegraphics[scale=0.98]{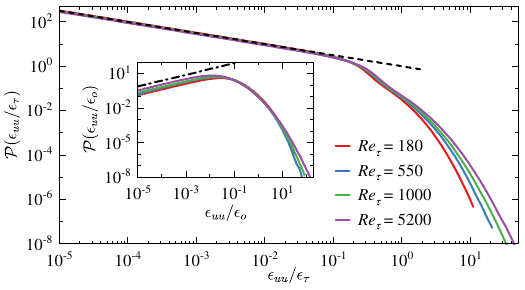}
\caption{Probability densities of $\epsilon_{uu}/\epsilon_\tau$ at the wall compared with those of $\epsilon_{uu}/\epsilon_o$ at the center (inset). Dashed line represents a slope of $-1/2$, dash-dotted line for slope of $1/2$. Here, $\epsilon_\tau$ is the wall dissipation (defined below Eq.~(2)) and $\epsilon_o$ is the scale of outer dissipation (defined just below Eq.~(5)). Data for $Re_\tau=180$ and $550$ are ours (see Supplemental Material \cite{supply}), and for $1000$ and $5200$ are from Ref.~\citep{LM2015}. }\label{fig:pdf_diss_uu}
\end{figure}

An important quantitative detail of the patterns is revealed by the probability density functions (PDFs) of $\epsilon_{uu}$ in Fig.~\ref{fig:pdf_diss_uu}. In particular, the well-collapsed left tails have a distinct slope of $-1/2$ on the wall and $1/2$ on the center (inset). These trends indicate a divergent PDF for zero dissipation at the wall and a vanishing PDF for the center. Both slopes have recently been reported in HIT for dissipation defined with different derivative terms \cite{Gotoh2023_PRL_diss_pdf, Gotoh2025_PRL_diss_pdf}. The oppositely signed slopes for the wall and the centerplane reflect different dynamics (also indicated by Fig.~\ref{fig:diss_snapshots}): they show that the standard energy cascade and the intermittency characteristic of the centerplane are strongly shaped at the wall by the shear, clearly amplifying the low-dissipation events (possibly also high-amplitude events).
They also highlight the role of the wall in shaping dissipation dynamics, leading to different scaling behaviors with changing distance from the wall, as stated toward the end of the Introduction. We now set out to establish these changes.

%A ``shoulder'' (at $\epsilon_{uu}^+\approx0.5$, absent in HIT) separating the left power-law and right stretched-exponential tails also reflects the influence of organized motions, possibly streaks, which delay the departure from the $-1/2$ regime.

% \begin{figure*}
% \includegraphics[width=1.0\textwidth]{figures/diss_uu_distribution_and_scaling.pdf}
% \caption{\label{fig:diss_uu_Re} Streamwise dissipation moments. (a) Normalized dissipation varying with the wall-normal distance.
% (b) $Re_\tau$-scaling of dissipation moments in the outer flow. Predictions of (\ref{eq:diss_outer_power}) are represented by lines, with solid from (\ref{eq:Sreeni-Yakhot_model}) and dashed from (\ref{eq:SL_model}). For clarity, data at different wall-normal positions are vertically shifted to align with the center values, without affecting the validation of scaling slopes.
% (c) $Re_\tau$-scaling of  dissipation moments at the wall. Dash-dotted lines indicate (\ref{eq:diss_wall_Townsend}), while solid lines indicate (\ref{eq:diss_wall_defect}). In both (b) and (c), colors denote different orders with arrows depicting increasing order for integer $p=1-4$. Data are the same as in Fig.\ref{fig:pdf_diss_uu}.
% The parameter values for the theoretical lines in (b) and (c) are listed in Table \ref{tab:parameters}.  }
% \end{figure*}

% \begin{figure}[b]
% \includegraphics[scale=1.0]{figures/diss_uu_distribution.pdf}
% \caption{\label{fig:diss_uu_distribution} Normalized dissipation $\langle\epsilon_{uu}\rangle/\epsilon_o$ varying with the wall-normal distance for different $Re_\tau$'s. The edges of the colored backgrounds are explored further in the text. Data are the same as in Fig.2.}
% \end{figure}

\begin{figure*}
\includegraphics[width=1.0\textwidth]{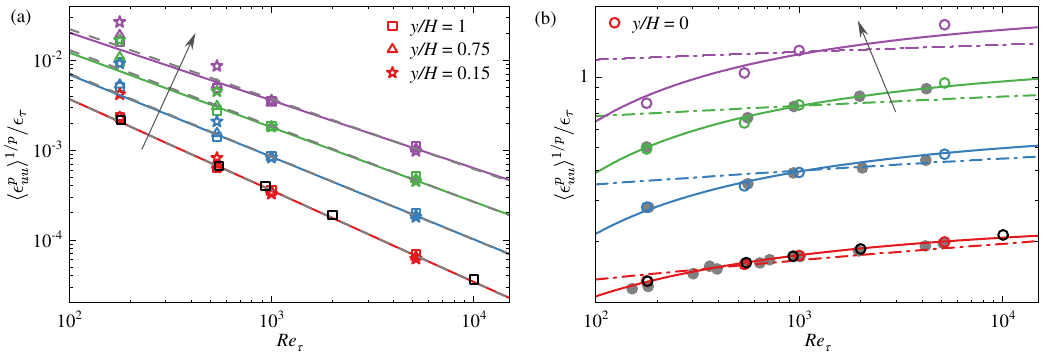}
\caption{\label{fig:diss_uu_Re} Streamwise dissipation moments. (a) $Re_\tau$-scaling of dissipation moments in the outer flow. Predictions of (\ref{eq:diss_outer_power}) are represented by lines, with solid from (\ref{eq:Sreeni-Yakhot_model}) and dashed from (\ref{eq:SL_model}). For clarity, data at different wall-normal positions are vertically shifted to align with the center values, without affecting the validation of scaling slopes.
(b) $Re_\tau$-scaling of  dissipation moments at the wall. Dash-dotted lines indicate (\ref{eq:diss_wall_Townsend}), while solid lines indicate (\ref{eq:diss_wall_defect}). In both panels, colors denote different orders with arrows depicting increasing order for integer $p=1-4$.
Gray symbols are data from Ref.\ \cite{Cheng2022_JFM_diss_wall}, black symbols are from Refs.\ \cite{Hoyas2008_DNS, Hoyas2022_DNS}, while others are the same as in Fig.~\ref{fig:pdf_diss_uu}.
The parameter values for the theoretical lines are listed in Table \ref{tab:parameters}.  }
\end{figure*}

\textit{Scaling in the outer region} --- We follow the framework developed for HIT \cite{Sreeni&Yakhot2021_PRF_scaling, Yakhot1998_PRE_pdf, Yakhot2006_pdf, Schumacher2007_NJP} and connect dissipation and inertial range dynamics. The derivation is given in detail in the Supplemental Material \cite{supply}, but a few steps are summarized here. By applying the point-splitting procedure to local velocity gradients \cite{Landau1987_fluid}, one has $\partial_x u \approx \delta_\eta u /\eta$. Extending the structure function scaling to both ends up to the dissipation and integral scales, together with the simplification of the exact equation for $p$th-order longitudinal structure functions \cite{Yakhot2001_PRE, Hill2001_JFM_all_orders} in the inertial range, one obtains the dissipation moments via the scaling of structure functions (equation (24) in Ref. \cite{Sreeni&Yakhot2021_PRF_scaling}), as
\begin{equation}
\label{eq:diss_Re_HIT_pre}
\frac{\left\langle\epsilon^p\right\rangle^{1/p}}{\sigma_L^{3}/L}
\sim Re_L^{\gamma_p}.
\end{equation}
Here, $\sigma_L=\sqrt{\langle (\delta u_L)^2\rangle}$ is the velocity difference over the large scale $L$, $Re_L=L \sigma_L /\nu$ is the corresponding Reynolds number, and the scaling exponent corrections $\gamma_p$ are expressed in terms of $\zeta_p$ as
\begin{equation}
\label{eq:gamma_p}
    \gamma_p = 1 + \frac{\zeta_{4p}}{p(\zeta_{4p} - \zeta_{4p+1} -1)}.
\end{equation}

Moreover, to apply (\ref{eq:diss_Re_HIT_pre}) to channels, two points deserve further clarification. First, according to the definitions, the large-scale $L$ of HIT corresponds naturally to the flow thickness $H$ of the channel. In HIT, $\sigma_L$ is the fluctuation intensity; for wall flows, since the mean velocity $U$ is not the right variable, $\sigma_L$ is chosen as $k=\sqrt{\langle uu \rangle}$, or as $u_\tau$, which is the commonly used metric for the fluctuation intensity in wall flows \citep{Hoyas2022_DNS}.
 Therefore, replacing $\sigma_L$ by $u_\tau$ and $L$ by $H$, (\ref{eq:diss_Re_HIT_pre}) reads as
\begin{equation}
\label{eq:diss_Re_HIT}
    \langle \epsilon_{}^{p} \rangle ^\frac{1}{p}/\epsilon_{o} \sim Re_\tau^{\gamma_p},
\end{equation}
where $\epsilon_{o}=u^3_\tau/H$.
Note that, compared to $\epsilon_\tau=u^4_\tau/\nu$ defined on the wall, $\epsilon_{o}$ is a more reasonable normalization for dissipation moments in the outer flow, due to the vanishing viscous effect. Also, since $\epsilon_o/\epsilon_\tau=Re^{-1}_\tau$ by definition, (5) immediately leads to (1).

It should be said that the outer $k$ profile is usually viewed as $Re$-independent when normalized by $u_\tau$ \citep{Nagib2024_PoF,Hultmark2012_PRL_Pipe, Ono2023_JFM_CSrelated}. This is reminiscent of the Townsend's outer similarity \citep{Townsend1976_AEM, book_Voyage_turb} established for mean velocity. However, one may surmise that (1) might change somewhat if there exists a weak Re-dependence of $k / u_\tau$. To clarify this, in the Supplemental Material \citep{supply}, we collect the DNS and experimental data for $Re_\tau$ reaching $5\times10^4$. It turns out that $k / u_\tau$ saturates at a constant for $Re_\tau$ above about 1000, thus implying that the choices of $\sigma_L=u_\tau$ or $\sigma_L=k$ are equivalent for large $Re_\tau$, with no change of (1). For moderate $Re_\tau$($\lesssim1000$), a weak dependence $k / u_\tau \sim Re^{0.028}_\tau$ does indeed occur. Yet, compared to the scaling exponent $-1+\gamma_p$ in (1), the additional correction is minor (at most about 9\%). Hence, the influence of choosing $\sigma_L=k$ is not essential, and the main conclusion here remains unchanged.

Second, the exponent $\gamma_p$ in (3) or (5) corresponds to an intermittent correction to the zeroth law \cite{Sreeni1984_PoF_diss, Vassilicos2015_ARFM_dissipation} and its generalization to higher order $p$'s. To see this, first note that K41's $ \zeta_{p}$ of $p/3$ yields $\gamma_p=0$ in \eqref{eq:gamma_p}. Consequently, $\langle\epsilon^p\rangle {^\frac{1}{p}}\sim \epsilon_o$, or $\langle\epsilon^p\rangle {^\frac{1}{p}}/\epsilon_\tau \propto Re^{-1}_\tau$. The $Re^{-1}_\tau$ scaling applies in a decent fashion for $p=1$, as shown in Fig.~\ref{fig:diss_uu_Re}a.
%Fig.~\ref{fig:diss_uu_distribution} by the data collapse of normalized dissipation profiles at different Reynolds numbers.
For higher $p$, the data differ increasingly from $Re^{-1}_\tau$ because of the increasing deviation of $ \zeta_{p}$ from $p/3$. Thus, a specific anomalous scaling model of $ \zeta_{p}$ is required to characterize the dissipation intermittency.

%as or equivalently $\langle\epsilon_{uu}^+\rangle Re_\tau = C_\epsilon^\prime$ with the constants $C_\epsilon$ and $C_\epsilon^\prime$ independent of Reynolds numbers.
%Fig. \ref{fig:Rek_and_diss_y}(b) show the distribution of normalized mean turbulent dissipation with varying wall-normal distance, which is collapse in core layer $y/H \gtrsim 0.5$ and even the whole outer region including the log layer $y/H \gtrsim 0.15$ once excluded the low-Reynolds-number effects ($Re_\tau=180$). This result indicates the zeroth law, as a cornerstone of turbulence theory \cite{Frisch_turb, Vassilicos2015_ARFM_dissipation}, together with theories built upon it, can then be extended to wall turbulence in regions not too distant from the wall.

Despite the close agreement among most intermittency models \cite{p-model, SL94, Frisch_turb, Sreeni&Yakhot2021_PRF_scaling} for low-order exponents, recent findings based on very high Reynolds number HIT simulations suggest the saturation of $\zeta_p$ when $p \to \infty$ \cite{Iyer2020_PRF_saturation, Sreeni&Yakhot2021_PRF_scaling, Buaria2023_PRL_saturation}. We thus adopt the particular model of Ref.\ \cite{Sreeni&Yakhot2021_PRF_scaling} that captures this saturation. That is,
\begin{equation}
\label{eq:Sreeni-Yakhot_model}
    \zeta_{p} = 7.3/(19/p+1).
\end{equation}
Taking (\ref{eq:gamma_p}) and (\ref{eq:Sreeni-Yakhot_model}) as valid, the prediction of (\ref{eq:diss_outer_power}) is shown in Fig.~\ref{fig:diss_uu_Re}a, in satisfactory agreement with the data for orders $p=1$ to 4. Note that three wall-normal positions are collected here from the log layer to the centerline (see Fig.~3), all obeying prediction (\ref{eq:diss_outer_power}). This universality indicates a minor effect of the mean shear on dissipation intermittency for the outer flow. Deviations in the data at $y/H=0.15$ at $Re_\tau=180$, also observed in earlier work \cite{Hamlington2012_JFM_diss, Schumacher2014_PNAS_universality}, are believed to be due to the low $Re$ effects with undeveloped inertial range \citep{DCS2025_JFM_circulation}.
The parameters used for the verification of (1) are summarized in Table I. In the Supplemental Material \cite{supply}, data sources and convergence of moments are presented in detail.

\begin{table}[b]
\begin{ruledtabular}
\begin{tabular}{ccccccc}
    Region & model & parameter & $p=1$ & $p=2$ & $p=3$ & $p=4$ \\
    \hline
    \multirow{2}{*}{Outer}  & {Eqs.~(1,~4,~6)}& $\gamma_p$ & -0.017 & 0.084 & 0.172 & 0.247 \\ \cline{2-7}
                           % &                       & $C_p$ & 0.40 & 0.47 & 0.55 & 0.65 \\ \cline{2-7}
                            & {Eqs.~(1,~4,~7)}& $\gamma_p$ & -0.017 & 0.077 & 0.156 & 0.222 \\
                            %&                       & $C_p$ & 0.40 & 0.50 & 0.63 & 0.80 \\
                            \hline
    \multirow{6}{*}{Wall}   & \multirow{4}{*}{Eq.~(2)}& $\alpha_{p,uu}$ & 0.25 & 0.63 & 1.22 & 2.02 \\
                            &                         & $\beta_{p,uu}$ & 0.42 & 1.31 & 2.62 & 4.34 \\
                            &                         & $\alpha_{p,ww}$ & {0.13} & {0.48} & {1.15} & {2.20} \\
                            &                         & $\beta_{p,ww}$ & {0.31} & {1.23} & {2.93} & {5.25} \\
                            \cline{2-7}
                            & \multirow{2}{*}{Eq.~(8)}& $A_p$ & 0.08 & 0.25 & 0.54 & 1.00 \\
                            &                         & $B_p$ & 0.01 & 0.02 & 0.03 & 0.04 \\
\end{tabular}
\end{ruledtabular}
\caption{\label{tab:parameters}%
{Parameter values for theoretical lines used in Fig.~\ref{fig:diss_uu_Re} and Fig.~\ref{fig:diss_ww_Re}. Note that for (1), values of $C_p$ are not shown here because Fig.~\ref{fig:diss_uu_Re}(a) verifies only the scaling exponent $\gamma_p$. Also note that $\gamma_p$ in Fig.~\ref{fig:diss_uu_Re}(a) and Fig.~\ref{fig:diss_ww_Re}(a) are the same values. }}
\end{table}

\begin{figure*}
\includegraphics[width=1.0\textwidth]{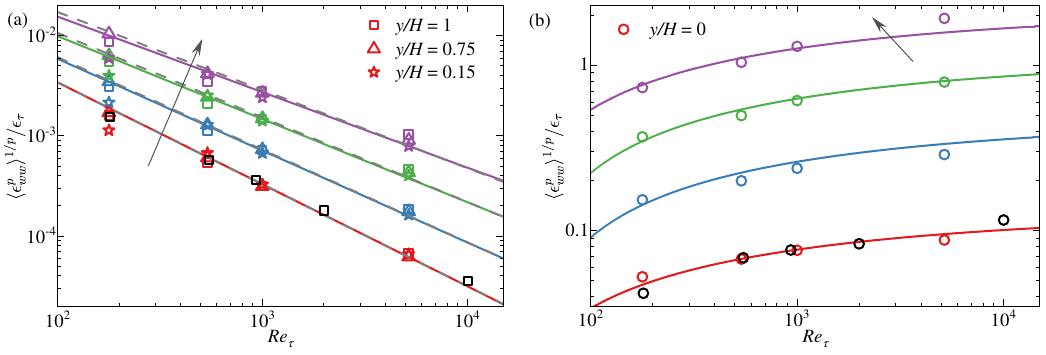}
\caption{\label{fig:diss_ww_Re}
Spanwise dissipation moments. Symbols and lines are the same as in Fig.\ \ref{fig:diss_uu_Re}. There are fewer data points (not reported in the literature) compared to the streamwise components. }
\end{figure*}

To show the robustness of (\ref{eq:diss_outer_power}), the log-Poisson model \cite{SL94} is now considered as an alternative to (\ref{eq:Sreeni-Yakhot_model}). That is,
\begin{equation}
\label{eq:SL_model}
    \zeta_{p} = p/9+2-2(2/3)^{p/3}.
\end{equation}
This model also adheres closely to the data (Fig.~\ref{fig:diss_uu_Re}a). However, as $p \to \infty$, it yields $\gamma_\infty=3/5$ in contrast to $\gamma_\infty=1$ arising from (\ref{eq:Sreeni-Yakhot_model}). These two models yield different results for large $p$ but not for the orders considered here.

\textit{Scaling at the wall --- }Dissipation exhibits fundamentally different behaviors on the wall.
As shown in Fig.~\ref{fig:diss_uu_Re}b, wall dissipation increases with $Re_\tau$, a trend that is in excellent agreement with the bounded defect law given by (\ref{eq:diss_wall_defect}). The basis of (\ref{eq:diss_wall_defect}) for $p=1$ has been discussed in \cite{CS21}, according to which the energy production limits the dissipation near the wall, which itself is bounded by a finite value, approached by a $-1/4$-power in $Re_\tau$. %The dependence on $Re^{-1/4}_\tau$ is thought to be the influence of outer Kolmogorov eddies.
Several consequences of this result have been deduced---e.g., the bounded behavior of the intensity of pressure fluctuation \cite{CS22}, the wall-normal profiles of velocity variance \citep{CS23} and its higher-order moments \citep{CS25}. There is also independent evidence of this behavior \cite{Ono2023_JFM_CSrelated, Yao2023_JFM_CSrelated, Monkewitz2022_JFM_CSrelated, Pirozzoli2024_JFM_CSrelated, Nagib2024_PoF}.

Equation (2) is an extension to higher orders of $p$ of this known behavior for $p=1$, via a linear $p$-norm Gaussian process. We have performed this extension for moments of velocity fluctuations in Ref.~\cite{CS25}. The novelty here is that we confirm its applicability to wall dissipation moments.
The order-dependent coefficients $\alpha_p$ are found to increase with $p$, which is consistent with a Gaussian-related random process (see Supplemental Material \citep{supply} for more discussion).

Also shown in Fig.~\ref{fig:diss_uu_Re}b as dashed dotted lines are the logarithmic growth prediction according to the attached-eddy model \cite{Townsend1976_AEM, Marusic2010_Science}, which states that
\begin{equation}
\label{eq:diss_wall_Townsend}
    \langle \epsilon_{}^{p} \rangle ^\frac{1}{p}/\epsilon_{\tau} =  A_p + B_p \ln Re_\tau.
\end{equation}
Here, $A_p$ and $B_p$ are constants independent of $Re_\tau$ but dependent on $p$. They are listed in Table \ref{tab:parameters} along with $\gamma_p$ of (1) and $\alpha_p$ and $\beta_p$ of (2).  The comparison of (\ref{eq:diss_wall_Townsend}) with the data, given in Ref.~\cite{Cheng2022_JFM_diss_wall} and reproduced here in Fig.~\ref{fig:diss_uu_Re}b for the same parameters, shows larger deviations than (\ref{eq:diss_wall_defect}), based on bounded dissipation.

In addition to the evidence from $\epsilon_{uu}$, the spanwise component of the dissipation, $\epsilon_{ww}=\nu\partial_j w \partial_j w$, is shown to exhibit the same Reynolds number scaling (Fig.\ref{fig:diss_ww_Re}) as the streamwise component, both in the outer region and at the wall.
This consistency reaffirms the scaling of Eqs.\ (1) and (2), as one might have expected.
We also note that no prediction of spanwise wall dissipation moments has been given before. The dissipation of wall-normal velocity $\epsilon_{vv}=\nu\partial_j v \partial_j v$ also satisfies the same law in the outer region, but it is zero on the wall (which means that $\alpha_p$ and $\beta_p$ in (\ref{eq:diss_wall_defect}) are zero because of the no-slip condition). They are therefore not shown here.

\textit{Discussion} --- We have examined here the boundary effects on dissipation intermittency, and obtained two fundamentally distinct Reynolds number scalings in turbulent channel flows. For the outer region where the mean shear is weak, we have identified a universal power law that extends the inertial-range intermittency to the dissipative range. Close to the wall where the mean shear is strong, the bounded defect law discussed here implies a non-intermittent character of dissipative events on the wall. Note that the mean dissipation at the wall remains to be O($Re_\tau)$ larger than that in the bulk. We point to three of several consequences of the present findings.

(\romannumeral1)  It is frequently argued (see, e.g., Refs.~[5], [6] and [14]) that intermittency requires increasingly finer grid resolution of DNS for resolving high-order moments of $\epsilon$ at larger Reynolds numbers. It may be thought that this consideration might apply for the dissipation near the channel walls as well \cite{Yang2021_PRF_grid_resolution}. However, the present result (see (2)) shows that all finite-high-order moments of $\epsilon$ are bounded, so that the DNS resolution could remain on the order ($\nu/u_\tau$) even at infinite Reynolds numbers. In contrast, (\ref{eq:diss_wall_Townsend}) requires a continually increasing grid resolution to resolve high-order wall-dissipation events as Reynolds number grows.

%For the numerical grid resolution needed for the DNS of wall flows when $Re_\tau \rightarrow \infty$, the number of cells in the grid scaled by the viscous units ($\nu/u_\tau$) is saturated for obtaining moments at any fixed order $p$, as they are bounded, see (\ref{eq:diss_wall_defect}), in contrast to  (\ref{eq:diss_wall_Townsend}), which requires an increasingly finer grid to resolve higher Reynolds number \cite{Yang2021_PRF_grid_resolution}.

%\begin{equation}
%\label{eq:diss_wall_CS}
%    \langle \epsilon \rangle %/\epsilon_{\tau} =  \alpha_1 - %\beta_1  Re_\tau^{-1/4}.
%\end{equation}

%It remains $Re$-dependent in non-saturating log-Poisson model:
%\begin{equation}
%\label{eq:diss_infty_norm_SL}
%    \lim_{p \to \infty} \langle \epsilon^{p}\rangle ^\frac{1}{p}/ \epsilon_{k}  \propto  Re_\tau^{3/5}.
%\end{equation}

%\begin{equation}
%\label{eq:diss_infty_norm_SY}
% \lim_{p \to \infty} \langle %\epsilon^{p}\rangle ^\frac{1}{p}/ %\epsilon_{k} \propto  Re{}_\tau;
%\end{equation}

(\romannumeral2)  The existence of outer flow intermittency, reflected by the non-zero exponent $\gamma_p$, and the absence of intermittency at the wall, as seen by a common exponent for all $p$ in (\ref{eq:diss_wall_defect}), is consistent with our recent findings on wall-parallel circulation \citep{DCS2025_JFM_circulation}. In fact, in the region where (\ref{eq:diss_outer_power}) applies, the intermittency of circulation (through the integration of $u$ and $w$ along closed Eulerian loops) takes the form of universal bifractality as in HIT \cite{Iyer2019_PRX_circulation}. However, close to the wall, the bifractality simplifies to a unifractal, implying that the circulation there is not intermittent. Equations (\ref{eq:diss_outer_power}) and (\ref{eq:diss_wall_defect}) support this picture.

(\romannumeral3) Finally, due to the concavity of $\zeta_p$ \cite{Frisch_turb}, one has $\gamma_p < 1$. Thus, for finite order $p$, (\ref{eq:diss_outer_power}) implies a negative scaling, hence vanishingly small dissipation in the bulk compared to that
on the wall, for very large Reynolds numbers.
%We know that the zeroth law does not hold for the bulk of the channel flow (i.e., a suitably normalized mean dissipation rate decreases with $Re_\tau$ as a weak power).
This means that away from the wall, the flow does not generate sufficiently strong singularities.
However, the extreme nature of dissipative events on the wall (compared to those in the bulk of the channel) suggests that walls may generate the necessary strength of singularities. This brings us to the broader possibility that the Onsager-type singularities \cite{Onsager1949_conjecture, Eyink2024_JFM_Onsager} might be generated only near the wall.

%We also note that, for normal scaling of velocity structure functions with $\zeta_p = p\zeta$ ($\zeta=1/3$ for K41 but not necessarily so in real turbulence \cite{Sreeni&Yakhot2021_PRF_scaling}), the relation (\ref{eq:gamma_p}) yields $\gamma_p = (1-3\zeta)/(\zeta+1)$, which is $p$-independent---just as the given scaling $-1/4$ of wall dissipation by Eq.  This remind us a recent observation regarding the near-wall non-fractal or space-filling statistics of velocity circulations \cite{DCS2025_JFM_circulation}, raising a intriguing possibility that wall dissipation, to some extent, behaves akin to a non-intermittent process and is related to near-wall non-intermittent eddies.

%Our findings bridge phenomenological observations of turbulence intermittency with mathematical questions concerning the singularities of the Navier-Stokes equations. In particular, they offer new information both in the limit of $Re \to \infty$ and moments orders $p \to \infty$, and contribute to the understanding of how turbulent dissipation and boundary effects intertwine.

%aknowledgements
We gratefully acknowledge the authors of \cite{LM2015} and the Johns Hopkins Turbulence Database \citep{JHU_database} for access to DNS data at $Re_\tau=1000$ and 5200. X. C. appreciates the support by the National Natural Science Foundation
of China, Nos.92252201, and the ``Fundamental Research Funds
for the Central Universities''.  KRS's research was supported by New York University.

\nocite{*}

\bibliography{main}% Produces the bibliography via BibTeX.

\end{document}

\preprint{APS/123-QED}

\title{Supplemental Material for \\
On how walls shape dissipation intermittency}

% \title{Supplemental Material}

 \author{Peng-Yu Duan}
 \author{Xi Chen}
% % \email{chenxi97@outlook.com}
 \affiliation{
  Key Laboratory of Fluid Mechanics of Ministry of Education, Beihang University (Beijing University of Aeronautics and Astronautics), Beijing 100191, PR China
 }
 \author{Katepalli R. Sreenivasan}
 \affiliation{%
  Tandon School of Engineering, Courant Institute of Mathematical Sciences, and Department of Physics, New York University, New York 10012, USA
 }%


\maketitle

\section{A. Direct numerical simulations}

In the main text, we consider cases with four friction
Reynolds numbers: $Re_\tau$ = 180, 550, 1000 and 5200.
For the higher Reynolds numbers $Re_\tau=1000$ and 5200, we have used the Johns Hopkins turbulence database \citep{JHU_database, LM2015}.
These data bases were generated by solving the incompressible Navier-Stokes equations using the pseudo-spectral (Fourier-Galerkin) method in $x$-$z$ planes, and a 7th-order B-spline collocation method in $y$-direction. The simulations correspond to a constant flow rate. Details on the settings and data can be found in \cite{LM2015}, but some flow parameters of general interest are listed in table \ref{tab:table1}.
\begin{table}
\begin{ruledtabular}
\begin{tabular}{cccccc}
    $Re_{\tau}$ & $L_x, L_z$ & $N_x\times N_z\times N_y %\times N_t
    $ & $\Delta x^+$ & $\Delta z^+$ & $\Delta y^+$ \\
    \hline
    180 & $4\pi,2\pi$ & $512\times 512\times 256 %\times 4500
    $ & 4.37 & 2.18 & 0.21--2.88 \\
    550 & $4\pi,2\pi$ & $1024\times 1024\times 512 %\times 2000
    $ & 6.61 & 3.30 & 0.17--4.1 \\
    1000 & $8\pi,3\pi$ & $2048\times 1536\times 512\times % 4000
    $ & 12.28 & 6.14 & 0.02--6.22 \\
    5200 & $8\pi,3\pi$ & $10240\times 7680\times 1536 %\times 11
    $ & 12.73 & 6.36 & 0.07--10.35 \\
\end{tabular}
\end{ruledtabular}
\caption{\label{tab:table1}%
Parameters of the simulations. $L_x,~L_z$ are the domain sizes in the streamwise and spanwise directions normalized by half channel height $H$, respectively. $N_x,~N_y,~N_z$ are the number of grid cells in the corresponding directions. $\Delta x^+$, $\Delta y^+$, and $\Delta z^+$ represent the grid spacings normalized by viscous units ($u_\tau,~\nu$). }
\end{table}

Data for $Re_\tau$ = 180 and 550 were obtained by us from direct numerical simulations of Navier-Stokes equations. We consider the incompressible equations
\begin{eqnarray}
    \partial_t \mathfrak{u}_i +\partial_j \mathfrak{u}_i \mathfrak{u}_j &=& -\partial_i \mathfrak{p} / \rho +\nu \partial_j \partial_j \mathfrak{u}_i~,
    \label{eq:ns_mom}
    \\
    \partial_i \mathfrak{u}_i &=& 0~,
\end{eqnarray}
where the indices $i=1,2,3$ indicate the streamwise $x$, wall-normal $y$, and spanwise $z$ directions, respectively. $\mathfrak{p}$ is the pressure, $\rho$ is the density, and $\mathfrak{u}_{i}$ denotes the instantaneous velocity components in the corresponding direction.
The fluctuating velocity $u_i$ defined as $u_i = \mathfrak{u}_{i} - \langle \mathfrak{u}_{i} \rangle$, where $\langle \cdot \rangle$ is the ensemble average, and $u_{1},u_{2},u_{3}=u,~v,w$ denote the streamwise, wall-normal, and spanwise fluctuating velocity components, respectively.

We use a finite difference code to solve the equations, which has been validated for channels in multiscale circulation investigation \citep{DCS2025_JFM_circulation}. Second-order central differences with very fine grids are employed for the discretization of spatial derivatives, and a second-order Runge-Kutta scheme is used for time advancement.
The computational domain of the channel, $L_x\times L_z$ (normalized by half the channel height $H$), is detailed in table \ref{tab:table1}. No-slip boundary conditions are imposed on both walls, and periodic boundary conditions are applied in streamwise and spanwise directions. The flows are driven by constant flow rates, with the initial condition of a Poiseuille velocity profile. The grid is uniform in both spanwise and streamwise directions, and is stretched from the wall to the centerline in the wall-normal direction, with the spacing shown in table \ref{tab:table1}.

\begin{figure*}
    \centering
    \includegraphics[width=0.97\linewidth]{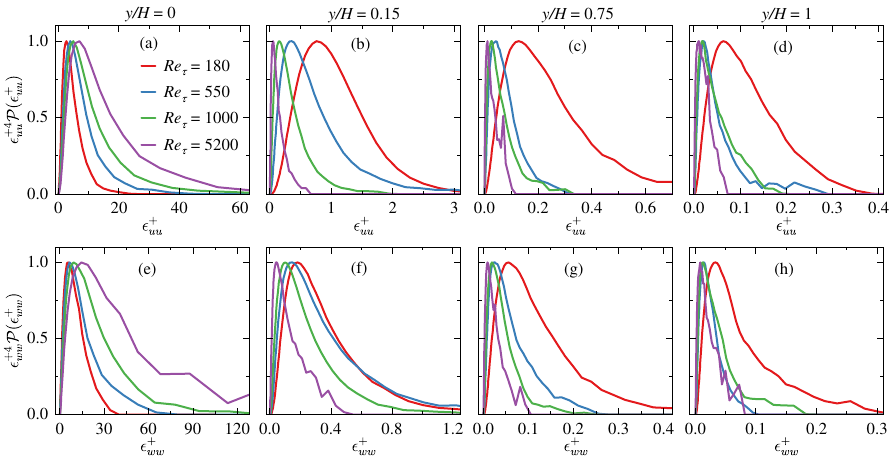}
    \caption{\label{fig:pre_pdfs} Pre-multiplied PDFs of wall-normalized dissipation rate, $\epsilon^+ = \epsilon / \epsilon_\tau$, for the 4th order moments. Here, $\epsilon_\tau$ is the mean flow dissipation at the wall as defined in the main text. (a-d): the streamwise components of dissipation, i.e., $\epsilon_{uu}=\nu \partial_j u \partial_j u$; (e-h): the spanwise components, i.e., $\epsilon_{ww}=\nu \partial_j w \partial_j w$.
    (a,e): results at the wall ($y/H=0$); (b,f) in the log layer ($y/H=0.15$); (c,g) in the core layer ($y/H=0.75$); (d,h) at channel center ($y/H=1$). }
\end{figure*}

\section{B. Outer scaling derivation}
We recapitulate for convenience the main steps of the derivation of Eqs.~(3-4) of the main text, from Refs.~\cite{Sreeni&Yakhot2021_PRF_scaling, Yakhot1998_PRE_pdf, Yakhot2006_pdf, Schumacher2007_NJP}.
Three different dissipation scales used in the derivations are explained here.
In contrast to the classical Kolmogorov scale  defined as $\eta = (\nu^3 / \langle \epsilon \rangle)^{1/4}$, the local dissipation scale $\tilde{\eta}$ is a time- and spatial-dependent length that satisfies $\tilde{\eta} \delta_{\tilde{\eta}} u  /\nu =1$, where $\delta_{\tilde{\eta}} u=u\left(x+\tilde{\eta}\right)-u\left(x\right)$. This $\tilde{\eta}$ will be used later to estimate the velocity derivative according to the point-splitting procedure. Another dissipation scale is an order-dependent $\eta_p$, defined as the crossover scale between the dissipative and inertial ranges of the $p$th-order longitudinal velocity structure function $S_p\left(r\right) \equiv \left\langle\left(\delta_ru\right)^p\right\rangle = \left\langle\left[u\left(x+r\right)-u\left(x\right)\right]^p\right\rangle$. In this way, the power law of $S_p(r)$ in the inertial range can be extended down to the scale of $S_p(\eta_p)$ \citep{Yakhot2006_pdf, Hamlington2012_JFM_diss}.

%It is also a statistical variable and can be related to the moments of $\tilde{\eta}$

Then, from the exact equations for $p$th-order longitudinal structure functions \cite{Hill2001_JFM_all_orders,Yakhot2001_PRE}, by taking $r \to \eta_p$ one obtains the estimate
\begin{equation}
\label{eq:S2p_S2p+1}
    \frac{S_{2p}\left(\eta_{2p}\right)}{\eta_{2p}}\approx\frac{S_{2p+1}\left(\eta_{2p}\right)}{\nu},
\end{equation}
which is Eq.(25) in Ref.\cite{Sreeni&Yakhot2021_PRF_scaling}. Recall the power law in the inertial range
\begin{equation}
\label{eq:Sp_power_law}
    S_{2p}\left(r\right)=A_{2p}\left(\frac{r}{L}\right)^{\zeta_{2p}};
\end{equation}
\begin{equation}
\label{eq:Sp_power_law2}
    S_{2p+1}\left(r\right)=A_{2p+1}\left(\frac{r}{L}\right)^{\zeta_{2p+1}}.
\end{equation}
Substitute (\ref{eq:Sp_power_law}) and (\ref{eq:Sp_power_law2}) into (\ref{eq:S2p_S2p+1}) with $r=\eta_{2p}$, and use the relation
$A_{2p+1}/A_{2p}\sim \sigma_L$ where $\sigma_L = \sqrt{\langle (\delta_L u)^2 \rangle}$, one obtains Eq.(26) of Ref.\cite{Sreeni&Yakhot2021_PRF_scaling}, i.e.
\begin{equation}
\label{eq:eta_L_scaling}
    \frac{\eta_{2p}}{L}\sim {Re_L}^\frac{1}{\zeta_{2p}-\zeta_{2p+1}-1}
\end{equation}
where $Re_L=L \sigma_L/\nu$.

%Since $\delta_L u$ is Gaussian ($L$ is the integral length scale), by setting $r=L$ in (\ref{eq:Sp_power_law}), we have $A_{2p}\sim \sigma_L^{2p}$, where $\sigma_L = \sqrt{\langle (\delta_L u)^2 \rangle}$.

In homogeneous and isotropic turbulence (HIT) or local HIT, dissipation can be estimated in terms of the one-dimensional derivative; and, through the point-splitting procedure, we have
\begin{equation}
\label{eq:epsilon_estimate}
\left\langle\epsilon^p\right\rangle^{1/p}
\sim\nu\left\langle\left(\frac{\partial u}{\partial x}\right)^{2p}\right\rangle^{1/p}
\sim\nu\left\langle\left(\frac{\delta_{\tilde{\eta}} u}{\tilde{\eta}}\right)^{2p}\right\rangle^{1/p}.
\end{equation}
According to the definition, $\delta_{\tilde{\eta}} u / \tilde{\eta}  = (\delta_{\tilde{\eta}} u)^2/\nu $, hence  (\ref{eq:epsilon_estimate}) is
\begin{equation}
\label{eq:epsilon_estimate2_AA}
\left\langle\epsilon^p\right\rangle^{1/p}
\sim \frac{\langle (\delta_{\tilde{\eta}}u)^{4p}\rangle^{1/p}}{\nu}.
\end{equation}
Further, with $\langle (\delta_{\tilde{\eta}} u)^p \rangle \sim \langle (\delta_{\eta_p} u)^p \rangle =S_p(\eta_p)$, one has $\langle (\delta_{\tilde{\eta}} u)^{4p} \rangle  \sim \langle (\delta_{\eta_{4p}} u)^{4p} \rangle = S_{4p}(\eta_{4p})$. Therefore, Eq.~(\ref{eq:epsilon_estimate2_AA}) is
\begin{equation}
\label{eq:epsilon_estimate2a}
\left\langle\epsilon^p\right\rangle^{1/p}
\sim \frac{[S_{4p}(\eta_{4p})]^{1/p}}{\nu}.
%\sim \frac{\sigma_L^{4}}{\nu} %\left(\frac{\eta_{4p}}{L}\right)^{\zeta_{4p} /p}
%\sim \frac{\sigma_L^{3}}{L} Re_L^{\gamma_p},
\end{equation}

Now, use $A_{2p}=(2p-1)!!\sigma^{2p}_{L}$ from (\ref{eq:Sp_power_law}) (considering Gaussian approximation of $\delta_L u$), together with (\ref{eq:eta_L_scaling}), from (\ref{eq:epsilon_estimate2a}) one obtains
\begin{equation}
\label{eq:epsilon_estimate2}
\left\langle\epsilon^p\right\rangle^{1/p}
\sim \frac{\sigma_L^{4}}{\nu} \left(\frac{\eta_{4p}}{L}\right)^{\zeta_{4p} /p}
\sim \frac{\sigma_L^{3}}{L} Re_L^{\gamma_p},
\end{equation}
where
\begin{equation}
\label{eq:dp}
    \gamma_p=1 + \frac{1}{p} \frac{\zeta_{4p}}{\zeta_{4p}-\zeta_{4p+1}-1}.
\end{equation}
Equations (S10-S11) are shown as Eqs.~(3-4) of the main text.

\section{C. Convergence of dissipation moments}
Figure \ref{fig:pre_pdfs} shows the pre-multiplied probability density functions (PDFs) for the 4th order of dissipation rate components (i.e. $p=4$, the highest order used in the current study) at different Reynolds numbers and different locations. The results show acceptable ``closure’' of the pre-multiplied PDFs in the sense that the integrands up to the fourth moment close reasonably well by the amount of available data.

\section{D. Reynolds number behavior of $k$}
As noted in the main text, one may choose $k= \sqrt{\langle uu \rangle}$ rather than $u_\tau$ as a surrogate of $\sigma_L$ in wall flows. Here, we argue that these two choices are consistent with each other because data shows that $k/u_\tau$ in the outer flow asymptote to a constant for large $Re_\tau$. In particular, figure S2(a) collects $k/u_\tau$ at the center line for $Re_\tau$ varying from $180$ to about $5\times10^4$. Data include DNS and CICLoPE pipe experiments that have been just presented in iTi2025 turbulence conference \citep{Hassan2025_iTi}.
Solid line marks the plateau of $k/u_\tau=0.91$, in excellent agreement with data for $Re_\tau$ above about 1000, thus justifying the equivalence between $\sigma_L=u_\tau$ and $\sigma_L=k$ in the outer wall flows.

For the moderate $Re_\tau$ range of $Re_\tau\lesssim1000$, Figure S2(b) a weak $Re$-dependence of $k/u_\tau \sim Re_\tau^s$ is apparent with $s\approx 0.028$. If choosing $\sigma_L = k$, Eq. (5) in the main text becomes
\begin{equation}
\label{eq:diss_outer_power_modify}
    \langle \epsilon_{}^{p} \rangle ^\frac{1}{p}/\epsilon_{o} \sim Re_\tau^{\gamma_p + s(\gamma_p+3)},
\end{equation}
and Eq.(1) is now
\begin{equation}
    \langle \epsilon_{}^{p} \rangle ^\frac{1}{p}/\epsilon_{\tau} \sim Re_\tau^{-1+\gamma_p + s(\gamma_p+3)}.
\label{eq:diss_outer_plus_power_modify}
\end{equation}
This new term, $s(\gamma_p+3)\approx 0.09$, would provides a minor correction to the exponent $-1+\gamma_p$ by about 9\%, if it is essential at all.

In the above, we have discussed the influence of choosing $\sigma_L=k$, which is restricted only for moderate $Re_\tau$. It should be emphasized that the conclusion and the consequences given in the main text remain unchanged.

\begin{figure}
\includegraphics[scale=1.2]{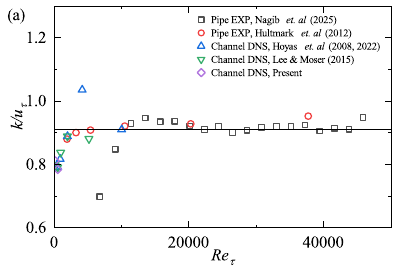}
\includegraphics[scale=1.2]{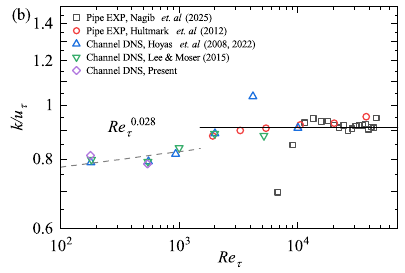}

\caption{(a) Data of $k/u_\tau$ at the channel and pipe centerline for $Re_\tau$ varying from 180 to about $5\times10^4$. (b) The same data as in (a) but with logarithmic axes. Dashed line indicates a power-law fit for $Re_\tau \leq 1000$, and solid lines indicate $k/u_\tau=0.91$. Experimental pipe data are from \cite{Hultmark2012_PRL_pipe, Hassan2025_iTi}; DNS channel data are from \cite{LM2015, Hoyas2008_DNS, Hoyas2022_DNS} and our present simulations.}

\label{fig:k_plus_Ret}
\end{figure}

\begin{figure}
    \centering
    \includegraphics[scale=1.0]{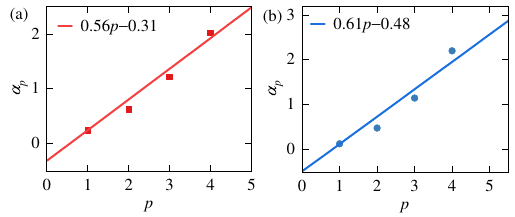}
    \caption{Values of $\alpha_p$ for (a) streamwise and (b) spanwise dissipation varying with order $p$. Lines correspond to the results of linear $p$-norm Gaussian.}
    \label{fig:LQNG}
\end{figure}

\section{E. Linear $p$-norm Gaussian process}

Fig.~\ref{fig:LQNG} shows the variation of $\alpha_p$ with increasing $p$ for the streamwise and spanwise dissipation moments. The linear $p$-norm Gaussian process enables an extension of the bounded defect-power law from $p=1$ to higher orders \cite{CS25}, which are marked by lines in the figure. {Overall, the lines are consistent with the trend of the fitting parameter $\alpha_p$, although slight difference observed.}

\section{F. Distribution of first-order dissipation}
Figure S4 shows the distribution of the first-order dissipation normalized by the outer characteristic dissipation $\epsilon_o = u_\tau^3 / H$.
Both the streamwise and spanwise components collapse well in the outer region for different Reynolds numbers, indicating an outer similarity for dissipation.

\begin{figure}
    \centering
    \includegraphics[scale=1.0]{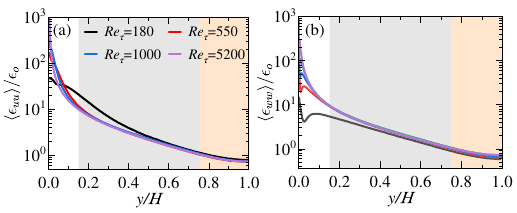}
    \caption{Normalized dissipation varying with the wall-normal distance for different $Re_\tau$'s for (a) the streamwise components and (b) the spanwise components. The edges of the
    colored backgrounds are explored further in the main text. Data
    are the same as in Fig. 2 of the main text.}
    
\label{fig:diss_1st_order}
\end{figure}

\bibliography{suppl}% Produces the bibliography via BibTeX.